\documentclass[lettersize,journal]{IEEEtran}
\usepackage{amsmath,amsfonts}
\usepackage{algorithmic}
\usepackage{array}
\usepackage{textcomp}
\usepackage{stfloats}
\usepackage{url}
\usepackage{moreverb}
\usepackage{verbatim}
\usepackage{graphicx}
\usepackage{subfigure}
\usepackage{bm}
\usepackage{amsthm}
\usepackage{cases}
\usepackage{amsmath}
\usepackage{amsthm,amsmath,amssymb}
\usepackage{mathrsfs}
\usepackage{amsmath} 
\usepackage{amssymb}
\usepackage[ruled]{algorithm2e}
\usepackage{caption}
\usepackage{booktabs}
\usepackage{multirow}
\usepackage{makecell}
\usepackage{tabu}
\usepackage{threeparttable} 
\usepackage{cite}

\def\BibTeX{{\rm B\kern-.05em{\sc i\kern-.025em b}\kern-.08em
    T\kern-.1667em\lower.7ex\hbox{E}\kern-.125emX}}
\usepackage{balance}
\begin{document}
\title{Resilient Clock Synchronization Architecture for Industrial Time-Sensitive Networking \\
}
\author{\IEEEauthorblockN{Yafei Sun, Qimin Xu, \emph{Member, IEEE}, Cailian Chen, \emph{Member, IEEE}, and Xinping Guan, \emph{Fellow, IEEE}}\\


}


\maketitle

\begin{abstract}
Time-Sensitive Networking (TSN) is a promising industrial Internet of Things technology. Clock synchronization provides unified time reference, which is critical to the deterministic communication of TSN. However, changes in internal network status and external work environments of devices both degrade practical synchronization performance. This paper proposes a temperature-resilient architecture considering delay asymmetry (TACD) to enhance the timing accuracy under the impacts of internal delay and external thermal changes. In TACD, an anti-delay-asymmetry method is developed, which employs a partial variational Bayesian algorithm to promote adaptability to non-stationary delay variation. An optimized skew estimator is further proposed, fusing the temperature skew model for ambiance perception with the traditional linear clock model to compensate for nonlinear error caused by temperature changes. Theoretical derivation of skew estimation lower bound proves the promotion of optimal accuracy after the fusion of clock models. Evaluations based on measured delay data demonstrate accuracy advantages regardless of internal or external influences.
\end{abstract}


\section{Introduction}

\IEEEPARstart{W}{ith} the popularization of Internet of Things technology, demand for deterministic and real-time communication has dramatically increased in various industries. In this context, an emerging set of Ethernet standards Time-Sensitive Networking (TSN), has become a potential communication mechanism in automotive and industrial automation \cite{8695835} since it guarantees the transmission of time-critical messages under bounded latency. Network mechanisms to ensure the deterministic communication capability in TSN are based on a unified time reference \cite{9194296}. Thus, clock synchronization determines the end-to-end transmission performance. IEEE 802.1AS standard, a specific profile of the common precise time protocol (PTP), is adopted for synchronization in TSN \cite{9121845}.

Multiple factors determine the effect of clock synchronization. For node devices that make up the TSN network, factors affecting synchronization come from the internal network connecting each device \cite{8064686, 9081999, 7234953, 7302563} and the external work environment of devices \cite{8125129, 6817598, GONG201788, 6836139}. Thus, the source of error can be divided into internal and external impacts, and considering them both \cite{9013260, 9214874, 9475453} is critical to improving synchronization accuracy. 

For internal impacts from network status, due to the two-way message exchange mechanism in 802.1AS, the bidirectional delay asymmetry in its propagation delay measurement is a significant error source \cite{9134100}. To this end, 802.1AS defined the data type and the correction scheme for delay asymmetry, but the estimation scheme has not been determined yet. Delay asymmetry is mainly caused by packet delay variation (PDV) between two-way communication \cite{8643987}. Although TSN mechanisms such as residence time measurement may attenuate PDV to some extent, it is inevitable due to limited implementation accuracy and additional introduced delays \cite{8697078, 7096944}. The errors caused by PDV accumulate with the increase of hops. Hence, reducing the delay asymmetry impact is necessary for promoting synchronization performance.

For coping with internal impacts, recent studies employ hybrid distribution models with better fitting performance to characterize propagation delays \cite{8064686, 9081999} due to the lack of prior knowledge about PDV. Nonetheless, there are large-scale and multi-type data in industrial scenarios. Various real-time streams dynamically change network status parameters \cite{8894896}, including load and background stream size, and can result in millisecond-level fluctuation of the PDV range \cite{9214874}. Moreover, the experiments using the network tester prove that dynamic network status can change statistical characteristics of delay asymmetries \cite{9589663}. Existing studies ignore the statistical dynamicity of delays and cannot accurately characterize their irregular distribution changes. Thus, it is essential to model PDV as non-stationary random in dynamic networks.

External environmental impacts such as humidity also affect synchronization accuracy. The traditional message exchange-based methods are based on the linear clock model, where the operating frequency is the slope value. However, external factors cause nonlinear variations in clock parameters that need to be estimated \cite{9475453}, where the temperature change is the main reason for clock instability \cite{GONG201788, 6817598}. Experiments prove that dynamic temperature reduces the effect of 802.1AS \cite{5981981}. In industrial scenarios such as steel plants, thermal changes from production are inevitable \cite{9214874}. Among them, automotive manufacturing standards stipulate that the onboard communication system must pass extreme thermal tests \cite{5981981}. The TSN synchronization method must deal with nonlinear errors from external environments. Hence, it is critical to consider the varying operating frequency caused by temperature changes.

For coping with external impacts, studies focusing on nonlinear error employ the frequency self-correction method based on the clock's temperature skew model \cite{6836139, 7839174}, which is verified by temperature chamber experiments \cite{8125129}. Nonetheless, considering that the dynamic communication network brings additional errors \cite{9214874}, synchronization faces dual challenges from network status and external environment, and any single clock model is challenging to meet TSN's actual needs. In this context, recent studies integrate the temperature skew model into message exchange-based methods \cite{GONG201788, 9013260, 6836139}. However, the inevitable temperature measurement error from limited sensor accuracy also causes serious estimation errors, leading to worse synchronization effects \cite{8125129, 7839174, 6836139}. Thus, simultaneously compensating for each clock model's defects is vital.

Considering the dual impacts of delay asymmetry in internal networks and nonlinear error from external environments mentioned above, this paper proposes a temperature-resilient architecture considering delay asymmetry (TACD) to tackle the problem of synchronization performance decline in TSN industrial scenarios. The main contributions are as follows:

\begin{itemize}
\item  An architecture TACD is built to promote synchronization accuracy for the precise industrial timing need. In TACD, the designed network communication and environment awareness phases estimate states of internal delay and external ambiance via message exchange and temperature measurement, respectively. Resulting estimates are analyzed jointly in the devised data fusion phase.

\item  An anti-delay-asymmetry method is developed to enhance adaptability to dynamic network status. Aiming at the statistical dynamicity of delay asymmetries, the method innovatively models PDVs as non-stationary. Then, a partial variational Bayesian algorithm is employed to iteratively estimate the parameters of the clock and PDV model, which improves estimation performance by accurate characterization for dynamic delay distribution.

\item  An optimized skew estimator is proposed to fuse resulting estimates from the network communication and environment awareness phases. For compensation of defects of the linear clock model and temperature skew model, the estimator is based on the limitation analysis of the two phases and employs Pareto optimization to balance the estimation bias and variance.
\end{itemize}

The rest of this paper is organized as follows. Section \uppercase\expandafter{\romannumeral2} discusses the related work. Section \uppercase\expandafter{\romannumeral3} describes TACD architecture. Section \uppercase\expandafter{\romannumeral4}, Section \uppercase\expandafter{\romannumeral5}, and Section \uppercase\expandafter{\romannumeral6} introduce the three synchronization phases. The optimal accuracy is derived in Section \uppercase\expandafter{\romannumeral7}. Performance evaluations are given in Section \uppercase\expandafter{\romannumeral8}, and Section \uppercase\expandafter{\romannumeral9} concludes.

\section{Related Work}

Studies on message exchange-based methods focus on reducing delay asymmetry impact. Communication delay consists of a fixed and a random part. For fixed parts, \cite{5605654} reveals that their two-way numerical relationship is essential for obtaining clock parameters. In this context, \cite{7302563} proposes prerequisite assumptions adopted by subsequent studies, in which the difference between two-way fixed delays is known. For random parts, considering their uncertainty, Gaussian and exponential distributions are used for modeling them in \cite{6582783}, and maximum likelihood estimations are adopted. \cite{7234953} models random delays with Gamma distributions, and a recursive estimation method is further developed. Recent studies employ hybrid distributions due to lacking knowledge of random delays in practice. Gamma and Gaussian mixture models (GMM) are adopted in \cite{8064686} and \cite{9081999}, respectively, where location-scale parameter estimations are used. The above studies focus on stochastic characteristics of delays without considering their non-stationary variations and clock nonlinear error. 

Among studies focusing on the clock's internal operation to reduce error from external environments, \cite{9475453} points out that ambient changes cause clock oscillator instability, resulting in time-varying skew. It is further indicated that temperature variation is the core factor \cite{6817598, 5981981}, and an approximating optimal estimation is proposed in \cite{6817598}. In this context, the clock's temperature skew model is utilized for frequency self-correction \cite{GONG201788, 6836139}, further verified by temperature chamber experiments in \cite{8125129}. Regression methods for model parameter estimation are described in detail \cite{8125129, 7839174}. However, the above studies did not consider the impact of network delays on model calibration and acquisition \cite{9214874}. Hence, synchronization performance decreases in dynamic networks.

Among recent studies simultaneously considering the impacts of network delay and environment temperature, \cite{9475453} employs Gaussian distribution to model random delay and proposes an extended Kalman filter-based method to track nonlinear varying clock parameters. The combination of linear programming for PDV and temperature compensation is developed in \cite{9013260}. \cite{6836139} introduces the temperature skew model into the message exchange-based method to achieve higher accuracy. A digital-twin-enabled method is proposed in \cite{9214874}, focusing on adapting to the unexpected PDV and temperature change. However, temperature measurement error from limited sensor accuracy causes a severe decline in synchronization effect \cite{8125129}. \cite{6836139} numerically analyzes the resulting estimation error and expounds its complex form, proving that the frequency self-correction degenerates to biased. \cite{7839174} further proposes that the resulting error is the same order of magnitude as the manufacturing tolerance of the oscillator. Thus, the issue of temperature measurement error is critical to be solved.

For the problem that synchronization performance reduces in industrial scenes, this paper establishes TACD architecture. For the PDV impact, the delay asymmetry is modeled as non-stationary to adapt to dynamic networks. For the temperature variation impact, TACD combines the temperature skew model to deal with thermal changes. The optimized estimator eliminates the shortcomings of each clock model through mutual compensation between synchronization phases.

\begin{figure*} [t]
\vspace{-0.22cm}
  \centering
{\includegraphics[width=13.70cm]{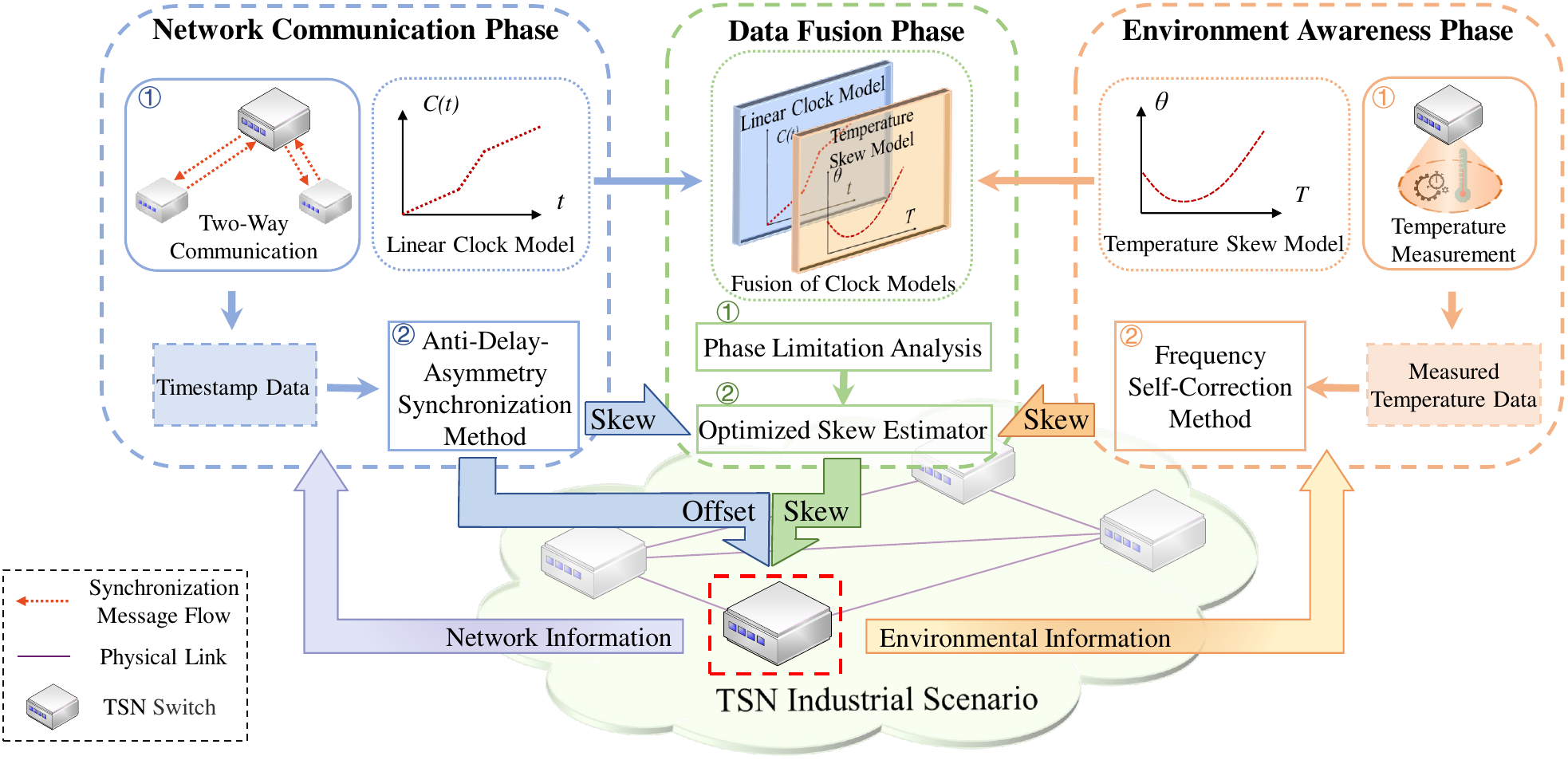}}
\vspace{-0.05cm}
  \caption{Architecture of TACD.}
\vspace{-0.17cm}
\end{figure*}

\section{TACD Architecture}

Clock parameters to be estimated include skew and offset. The clock output with deviation at time \emph{$t$} is expressed as:
\begin{equation}
\begin{split}
C(t)=\big(\theta(t)+1\big)\cdot t+\delta(t),
\end{split}
\end{equation}
where \emph{$C(t)$} is the practical time of the clock to be synchronized. \emph{$\theta(t)$} and \emph{$\delta(t)$} represent skew and offset. For coping with the clock parameter estimation problem, a TACD architecture is proposed, which contains three phases shown in Fig. 1. The network communication phase estimates clock parameters through collected network information. The environment awareness phase obtains clock skew using measured temperature. Based on estimates from the prior two phases, the data fusion phase estimates optimized skew. From the synchronization stages above, needed clock parameters are obtained. The steps are clarified below.

\subsubsection{Network Communication Phase}

This phase estimates clock parameters based on synchronization messages in the network. Firstly, timestamp data are obtained through two-way communication between TSN switches. Then, based on the linear clock model, skew and offset are estimated through the anti-delay-asymmetry synchronization method, where PDVs are modeled as non-stationary to adapt to dynamic network status. The detailed process is presented in Section \uppercase\expandafter{\romannumeral4}.

\subsubsection{Environment Awareness Phase}

This phase estimates the clock frequency deviation through the perception of external environments. Firstly, the temperature value of TSN switches is measured using the built-in sensor. Then, based on the ready-made temperature skew model, the clock skew is estimated using the frequency self-correction method without network communication. This phase considers the clock's internal operation to overcome the impact of work environments. The specific process is introduced in Section \uppercase\expandafter{\romannumeral5}.

\subsubsection{Data Fusion Phase}

This phase integrates the skew estimates from the previous two phases to obtain an optimized value. The limitation analysis for each phase is carried out first. Then the skew estimates based on various clock models are fused through Pareto optimization to balance the estimation bias and variance. The defects of each clock model can be compensated through the fusion of models. The specific process is introduced in Section \uppercase\expandafter{\romannumeral6}.

\section{Network Communication Phase}

This section introduces the network communication phase. Based on the linear clock model, a state-space model is established where PDVs are modeled as non-stationary. A partial variational Bayesian algorithm is then employed to estimate the model parameters of the clock and delay variation.

\subsection{Linear Clock Model}

The message exchange-based methods, including 802.1AS, are based on the linear clock model. The propagation delay is first estimated through periodic two-way communication. In this process of the \emph{$k$}-th period, the timestamps \emph{$t_{k}^{\mathsf{1}}$} and \emph{$t_{k}^{\mathsf{2}}$} are obtained through the first synchronization message transmission, while \emph{$t_{k}^{\mathsf{3}}$} and \emph{$t_{k}^{\mathsf{4}}$} are obtained through the second. Among them, \emph{$t_{k}^{\mathsf{1}}$} and \emph{$t_{k}^{\mathsf{4}}$} are instantaneous times of the clock to be synchronized, while the other two are that of the reference clock. The single-hop delay is estimated from the obtained timestamp data and is accumulated to calculate the delay of the entire link. The above process is modeled as \cite{7234953, 8064686}:
\begin{equation}
\begin{split}
t_{k}^{\mathsf{2}}=t_{k}^{\mathsf{1}}+d_{1}+w_{k}^{\mathsf{1}}+\delta_k, \\
t_{k}^{\mathsf{3}}=t_{k}^{\mathsf{4}}-d_{2}-w_{k}^{\mathsf{2}}+\delta_k,
\end{split}
\end{equation}
where \emph{d$_{1}$} and \emph{d$_{2}$} are fixed delays, and \emph{$w_{k}^{\mathsf{1}}$} and \emph{$w_{k}^{\mathsf{2}}$} are random delays in the two directions. $\delta_k$ is the offset in the \emph{$k$}-th period. 

The clock skew is caused by oscillator operation deviation from nominal frequency due to the equipment and external factors. The skew results in offset after accumulating over some time. Thus, the skew is regarded as the slope of the linear clock model and is described as the offset's change rate \cite{7234953, 6817598}. The estimated expression in the \emph{$k$}-th period is:
\begin{equation}
\begin{split}
\hat{\theta}^{\mathbb{L}}_k=\frac{\Delta\hat{\delta}_k}{\tau}=\frac{\hat{\delta}_k-\hat{\delta}_{k-1}}{t_{k}^{\mathsf{1}}-t_{k-1}^{\mathsf{1}}},
\end{split}
\end{equation}
where $\hat{\theta}^{\mathbb{L}}_{k}$ is the estimated skew based on the linear clock model, \emph{$\tau$} denotes the fixed synchronization period, and \emph{$\Delta\hat{\delta}_k$} is the estimated offset increment. In this way, clock parameters can be estimated by the message exchange-based method.

For fixed delays, a fundamental precondition for estimation is that their difference is known \cite{8643987, 7302563}, i.e., \emph{$d=d_{1}-d_{2}$}, where \emph{$d$} denotes the known fixed delay asymmetry. This universal precondition covers the case of equal unknown fixed delays in \cite{9475453, 8064686} and is employed in this paper. 

For random delays, their statistical parameter changes have yet to be considered. We model random delays as non-stationary to characterize their actual dynamicity, thus promoting adaptability to dynamic network status.

\subsection{Anti-Delay-Asymmetry Synchronization Method} 

Considering the clock operation mechanism, the first-order Gaussian Markov model describes this random variation process \cite{9475453}, where the skew expression at the \emph{$k$}-th period is:
\begin{equation}
\begin{split}
\theta_k=m\theta_{k-1}+u_k,
\end{split}
\end{equation}
where \emph{$m$} denotes the constant state transfer coefficient. \emph{$u_k$} is Gaussian distribution noise with mean 0 and variance \emph{$\sigma_{u}^2$}.

The clock offset is formed by the composition of the initial manufacturing error and the skew cumulative amount during operation, whose recursive state model is expressed as \cite{9475453}:
\begin{equation}
\begin{split}
\delta_k=\delta_{k-1}+\tau\theta_k.
\end{split}
\end{equation}

Models (4) and (5) are integrated into the clock parameters state equation to characterize skew and offset simultaneously:
\begin{equation}
\begin{split}
\mathbf{x}_k=\mathbf{Ax}_{k-1}+\mathbf{v}_k,
\end{split}
\end{equation}
where $\mathbf{x}_k=\begin{bmatrix}  \   \theta_k    \ \  \\  \   \delta_k   \ \  \end{bmatrix}$, $\mathbf{A}=\begin{bmatrix}  \   m &    0   \ \  \\  \   m\tau  &    1 \ \  \end{bmatrix}$, and $\mathbf{v}_k=\begin{bmatrix}  \   u_k  \ \  \\  \   \tau u_k \ \  \end{bmatrix}$. State vector $\mathbf{x}_k$ contains clock parameters in the \emph{$k$}-th period, $\mathbf{A}$ denotes the state transition matrix, and $\mathbf{v}_k$ is the noise vector. Since $u_k$ confirms to Gaussian distribution, thus $\mathbf{v}_k$ is a Gaussian noise vector with mean vector $\bm{0}$ and covariance matrix \emph{$\mathbf{Q}_{v}=\begin{bmatrix}  \   \sigma_{u}^2  &  \tau \sigma_{u}^2  \ \  \\  \   \tau \sigma_{u}^2  &  \tau^2 \sigma_{u}^2 \ \  \end{bmatrix}$}.

In the message exchange-based method, the offset is estimated based on the linear clock model through the two-way communication process in (2) and is expressed as:
\begin{equation}
\begin{split}
\hat{\delta}^{\mathbb{L}}_k=\frac{(t_{k}^{\mathsf{2}}+t_{k}^{\mathsf{3}}-t_{k}^{\mathsf{1}}-t_{k}^{\mathsf{4}})-d}{2}+\frac{w_{k}^{\mathsf{2}}-w_{k}^{\mathsf{1}}}{2}.
\end{split}
\end{equation}

Similarly, the one-way message transmission between switches in two adjacent periods can be modeled in light of (2) to obtain the offset increment. According to characterization in (3), the estimated clock skew is expressed as \cite{7234953, 6817598}:         
\begin{equation}
\begin{split}
\hat{\theta}^{\mathbb{L}}_k=\frac{t_{k}^{\mathsf{2}}-t_{k-1}^{\mathsf{2}}-t_{k}^{\mathsf{1}}+t_{k-1}^{\mathsf{1}}}{\tau}+\frac{w_{k-1}^{\mathsf{1}}-w_{k}^{\mathsf{1}}}{\tau}.
\end{split}
\end{equation}

Therefore, the clock parameters observation equation composed of (7) and (8) is expressed as:
\begin{equation}
\begin{split}
\mathbf{z}_k=\mathbf{Hx}_k+\mathbf{n}_k,
\end{split}
\end{equation}
where $\mathbf{z}_k= \begin{bmatrix}  \   t_{k}^{\mathsf{2}}-t_{k-1}^{\mathsf{2}}-t_{k}^{\mathsf{1}}+t_{k-1}^{\mathsf{1}}   \ \  \\  \   t_{k}^{\mathsf{2}}+t_{k}^{\mathsf{3}}-t_{k}^{\mathsf{1}}-t_{k}^{\mathsf{4}}-d   \ \  \end{bmatrix}$, $\mathbf{H} = \begin{bmatrix}  \   \tau & 0   \ \  \\   \    0 & 2  \ \  \end{bmatrix}$, and $\mathbf{n}_k = \begin{bmatrix}  \   w_{k}^{\mathsf{1}}-w_{k-1}^{\mathsf{1}}   \ \  \\   \    w_{k}^{\mathsf{1}}-w_{k}^{\mathsf{2}}  \ \  \end{bmatrix}$. Measurement vector $\mathbf{z}_k$ contains obtained timestamp data and known fixed delay asymmetry, $\mathbf{H}$ is the measurement matrix, and $\mathbf{n}_k$ is the noise vector. 

Measurement noise vector $\mathbf{n}_k$ contains PDVs $w_{k}^{\mathsf{1}}-w_{k-1}^{\mathsf{1}}$ and $w_{k}^{\mathsf{1}}-w_{k}^{\mathsf{2}}$. Due to the lack of knowledge about random delays, GMMs are used to model PDV distributions. For the mean values, since delay variation is completely random under irregular network status changes, the probability of different trends is identical. Thus, the PDV mean values are zero. For the variances, since dynamic networks cause changes in statistical parameters and fluctuation ranges of delays, modeling PDV variances as time-varying can better reflect delay dynamicity. Thus, $\mathbf{n}_k$ is modeled as a non-stationary GMM vector, whose distribution is formulated as:
\begin{equation}
\begin{split}
p(\mathbf{n}_k)=\sum\nolimits_{i=1}^{N_{g}}a_k^i\mathcal{N}(\mathbf{n}_k|\mathbf{0},\mathbf{R}_{k}^i), \  \sum\nolimits_{i=1}^{N_{g}}a_k^i=1,
\end{split}
\end{equation}
where $N_{g}$ is the number of Gaussian components, and $a_{k}^i$ and $\mathbf{R}_{k}^i$ represent the weight and covariance matrix of the \emph{$i$}-th Gaussian component in the \emph{$k$}-th period, respectively. By updating noise vector parameters, the established state-space model can characterize dynamic network delay.

Unlike previous studies only estimating clock parameters, the proposed anti-delay-asymmetry synchronization method also updates PDV model parameters. However, when non-stationary GMMs are used to model PDVs, employing traditional variational Bayesian inference cannot maintain the posterior distribution conjugacy \cite{8933396}, making it challenging to jointly estimate the required state and noise. 

On this account, a partial variational Bayesian algorithm ensuring conjugacy \cite{8933396} is employed, where the state and noise are estimated alternately in each period. For clock parameters, based on the PDV model updated in the previous period, state posterior distribution in the current period is estimated via the Gaussian sum filter \cite{7523217} and is expressed as:
\begin{equation}
\begin{split}
p(\mathbf{x}_k|\mathbf{z}_{1:k})=\sum\nolimits_{i,j}\hat{\omega}_{k|k}^{ij}\mathcal{N}(\mathbf{x}_k|\hat{\mathbf{x}}_{k|k}^{ij},\mathbf{P}_{k|k}^{ij}),
\end{split}
\end{equation}
where $\hat{\omega}_{k|k}^{ij}$, $\hat{\mathbf{x}}_{k|k}^{ij}$, and $\mathbf{P}_{k|k}^{ij}$ are each component's weight, state estimate, and covariance matrix. $\mathbf{z}_{1:k}$ is the set of measurements up to and including the \emph{$k$}-th period. Through the state estimation, the required clock parameters are obtained.

For parameters in PDV model $\mathbf{n}_k$, their posterior distribution is updated by variational Bayesian inference. In this process, a discrete latent variable $\mathbf{y}_k$ is employed to characterize Gaussian component weights \cite{8933396}. The PDV model parameter set $\mathbf{\Psi}_k=\{{\mathbf{y}_k}, {\bm{a}_k}, {\mathbf{R}_k}\}$, where $\mathbf{y}_{k}=\{y_k^i\}_{i=1}^{N_{g}}$, $\bm{a}_{k}=\{a_k^i\}_{i=1}^{N_{g}}$ and $\mathbf{R}_{k}=\{\mathbf{R}_k^i\}_{i=1}^{N_{g}}$. The optimal solution $p^{*}(\mathbf{\Psi}_k)$ of the posterior distribution approximation is expressed as \cite{9167469}: 
\begin{equation} 
\begin{split}
{\rm log}\ p^{*}(\mathbf{\Theta}_k)={\rm E}_{\mathbf{\Psi}_k^{(\mathbf{-\Theta}_k)}}[{\rm log}\ p(\mathbf{\Psi}_k, \mathbf{z}_{1:k}|\hat{\mathbf{x}}_{k|k})]+c_{\mathbf{\Theta}_k},
\end{split}
\end{equation}
where $\mathbf{\Theta}_k$ is an arbitrary item in $\mathbf{\Psi}_k$, ${\mathbf{\Psi}_k^{(\mathbf{-\Theta}_k)}}$ refers to items in $\mathbf{\Psi}_k$ except for $\mathbf{\Theta}_k$, and $c_{\mathbf{\Theta}_k}$ is a constant concerning ${\mathbf{\Theta}_k}$. The distribution $p(\mathbf{\Psi}_k, \mathbf{z}_{1:k}|\hat{\mathbf{x}}_{k|k})$ can be decomposed as:
\begin{equation}
\begin{split}
\!p(\mathbf{\Psi}_k, \!\mathbf{z}_{1:k}|\hat{\mathbf{x}}_{k|k}\!)\!=\!p(\mathbf{z}_{k}|\mathbf{\Psi}_k, \!\hat{\mathbf{x}}_{k|k}\!)p(\mathbf{\Psi}_k|\mathbf{z}_{1:k-1}\!)p(\mathbf{z}_{1:k-1}\!),\!
\end{split}
\end{equation}
where the conditional likelihood distribution of $\mathbf{z}_{k}$ is given by:
\begin{equation}
\begin{split}
p(\mathbf{z}_{k}|\mathbf{\Psi}_k, \hat{\mathbf{x}}_{k|k})=\prod_{i=1}^{N_g}\mathcal{N}(\mathbf{z}_{k}|\mathbf{H}\hat{\mathbf{x}}_{k|k}, \mathbf{R}_k^i)^{y_k^i}.
\end{split}
\end{equation}

To guarantee form consistency during updating, given that the normal distribution covariance matrix and multinomial distribution have both conjugate prior distributions, the prediction distribution of PDV model parameters is expressed as \cite{8933396}:
\begin{equation}
\begin{split}
\!&p(\mathbf{\Psi}_k|\mathbf{z}_{1:k-1})\!=\!p(\mathbf{y}_k|\mathbf{z}_{1:k-1})p(\bm{a}_k|\mathbf{z}_{1:k-1})p(\mathbf{R}_k|\mathbf{z}_{1:k-1})\!\\
\!&=\prod_{i=1}^{N_g}(a_k^i)^{y_k^i}\mathcal{D}(\bm{a}_k|\hat{\bm{\chi}}_{k|k-1})\mathcal{IW}(\mathbf{R}^i_k|\hat{\upsilon}^i_{k|k-1}, \hat{\mathbf{V}}^i_{k|k-1}),\!
\end{split}
\end{equation}
where $\mathcal{D}(\bm{\chi})$ denotes a Dirichlet distribution with parameter vector $\bm{\chi}$, $\mathcal{IW}(\upsilon, \mathbf{V})$ is an inverse Wishart distribution with freedom degree ${\upsilon}$, and scale matrix $\mathbf{V}$. From (14) and (15), distribution $p(\mathbf{\Psi}_k, \mathbf{z}_{1:k}|\hat{\mathbf{x}}_{k|k})$ in (13) can be calculated, and the optimal solution (12) is further acquired. Model parameters of $\mathbf{n}_k$ are estimated by substituting $\mathbf{\Theta}_k$ with items in $\mathbf{\Psi}_k$. Then, the required PDV model is obtained. 

Through iterative estimation of clock parameters and PDV model parameters, whose calculation steps have been derived \cite{8933396}, the network communication phase can adapt to non-stationary delay asymmetry in dynamic TSN networks.

\section{Environment Awareness Phase}

In this phase, frequency self-correction is based on periodic ambient temperature measurements to estimate clock skew. The temperature skew model is obtained by regression operation of skew-temperature value pairs and is \cite{8125129, 7839174, 6836139}:
\begin{equation}
\begin{split}
\hat{\theta}^{\mathbb{T}}_{k}=\kappa(\tilde{T}_k-T_{0})^2+\theta_{0},
\end{split}
\end{equation}
where $\hat{\theta}^{\mathbb{T}}_{k}$ is the estimated clock skew based on the temperature skew model, \emph{$\tilde{T}_k$} is the measured temperature in the \emph{$k$}-th period, \emph{$T_{0}$} is the ideal temperature value, \emph{$\kappa$} denotes the sensitivity factor, and \emph{$\theta_{0}$} is the inherent clock skew. Various methods for regression operation have been discussed in detail in \cite{8125129, 7839174, 9214874}. Since this paper focuses on employing the ready-made model to compensate for the nonlinear error, it is assumed that model parameters have already been obtained through the calibration process in the early stage and the temperature skew model is known entirely throughout.

Limited sensor accuracy \cite{6836139, 8125129} and set sampling interval \cite{7839174} result in temperature measurement error, which is the leading cause of skew estimation error. The measured temperature value in the \emph{$k$}-th period is expressed as:
\begin{equation}
\begin{split}
\tilde{T}_k=T_k+\xi_k,
\end{split}
\end{equation}
where \emph{$T_k$} represents the actual temperature. \emph{$\xi_k$} is the measurement error, modeled as Gaussian distribution with mean 0 and variance \emph{$\sigma_{T}^2$} as in \cite{6836139}. Through the perception of ambiance, the frequency self-correction method can estimate the clock skew. Limitations of the environment awareness phase will be discussed in Section \uppercase\expandafter{\romannumeral6}.

\section{Data Fusion Phase}

This section introduces the data fusion phase. The limitation analysis for various synchronization phases is first carried out, based on which a skew estimator employing Pareto optimization to fuse the two clock models is then proposed.

\subsection{Limitation Analysis for Each Synchronization Phase}

In the previous phases, skew values are estimated based on different clock models. In the network communication phase, PDV randomness leads to inevitable errors so that the skew estimate contains actual skew $\theta_{k}$ and estimation error $e^{\mathbb{L}}_{k}$. Since modeling using GMMs and estimating using the Gaussian sum filter are unbiased \cite{7523217}, the skew estimate expectation ${\rm{E}}\{\hat{\theta}^{\mathbb{L}}_{k}\}$ equals the actual skew after the early stages. Thus, the bias $\mu^{\mathbb{L}}_{{k}}={\rm{E}}\{e^{\mathbb{L}}_{k}\}$ of the estimation error is $0$.

Given the covariance matrix $\hat{\mathbf{P}}_{k|k}$ of the posterior state estimation error of $\hat{\mathbf{x}}_{k|k}$, which characterizes skew and offset estimation performance, the variance $({\sigma}^{\mathbb{L}}_{k})^2$ of error $e^{\mathbb{L}}_{k}$ is equal to the first row, first column element of $\hat{\mathbf{P}}_{k|k}$, denoted as $\epsilon_k$. 

Further, the correlation of estimation error $e^{\mathbb{L}}_{k}$ is given as:
\begin{equation}
\begin{split}
{\rm{E}}\{(e^{\mathbb{L}}_{k})^2\}=({\sigma}^{\mathbb{L}}_{k})^2+({\rm{E}}\{e^{\mathbb{L}}_{k}\})^2=\epsilon_k.
\end{split}
\end{equation}

From this, although the Gaussian sum filter in the anti-delay-asymmetry synchronization method minimizes the mean square error of estimation, the error caused by the randomness of unexpected PDV is still inevitable, which limits the performance of the linear clock model-based estimation. 

In the environment awareness phase, inevitable temperature measurement errors degrade accuracy. Estimation error $e^{\mathbb{T}}_{k}$ of the frequency self-correction method is expressed based on the measurement model (17). The bias of $e^{\mathbb{T}}_{k}$ is further given as:
\begin{equation}
\begin{split}
&\mu^{\mathbb{T}}_{k}={\rm{E}}\{e^{\mathbb{T}}_{k}\}={\rm{E}}\{2\kappa(T_k-T_{0})\cdot \xi_k\}+\kappa{\rm{E}}\{\xi_{k}^2\}\\&=2\kappa(T_k-T_{0})\cdot {\rm{E}}\{\xi_k\}+\kappa\Big(\sigma_{T}^2+({\rm{E}}\{{\xi_k}\})^2\Big)=\kappa\sigma_{T}^2.
\end{split}
\end{equation}

From this, the temperature skew model-based estimation is biased, and the correlation of $e^{\mathbb{T}}_{k}$ is further calculated as:
\begin{equation}
\begin{split}
{\rm{E}}\{(e^{\mathbb{T}}_{k})^2\}
&=\kappa^2\Big(4\sigma_{T}^2(T_k-T_{0})^2+3\sigma_{T}^4\Big).
\end{split}
\end{equation}

From the above analysis, the network communication phase may cause a significant estimation variance, and the environment awareness phase may have a large estimation bias. Given that synchronization faces impacts from internal network status and external work environment simultaneously, any single clock model-based phase is challenging to meet TSN needs.

Considering that estimation errors caused by each synchronization phase are uncoupled, the combination of the two phases can compensate for defects of each clock model. Frequency self-correction can compensate for nonlinear errors in message exchange-based estimation. Method considering delay asymmetry can compensate for the impact of network PDV on the temperature skew model calibration \cite{9214874}. Therefore, the fusion of resulting estimates based on each clock model can achieve better synchronization performance.

\subsection{Optimized Clock Skew Estimator}

For combining the two phases to cope with internal and external impacts, an optimized skew estimator is proposed to overcome the limitations of each clock model. After obtaining skew values from respective phases, the estimate through linear fusion in the \emph{$k$}-th period is expressed as:
\begin{equation}
\begin{split}
\hat{\theta}^{\mathbb{F}}_{k}=\alpha_{k}\hat{\theta}^{\mathbb{L}}_{k}+\beta_{k}\hat{\theta}^{\mathbb{T}}_{k}=\theta_{k}+e^{\mathbb{F}}_{k}, \  \alpha_{k}+\beta_{k}=1,
\end{split}
\end{equation}
where $e^{\mathbb{F}}_{k}$ is the fusion-based estimation error. $\alpha_{k}$ and $\beta_{k}$ are fusion weight values, which should be chosen optimally. Given the limitation analysis above, the estimation error $e^{\mathbb{F}}_{k}$ can be expressed, and the bias of which is further calculated as:
\begin{equation}
\begin{split}
&\mu^{\mathbb{F}}_{k}={\rm{E}}\{e^{\mathbb{F}}_{k}\}={\rm{E}}\{\alpha_{k}(\theta_{k}+e^{\mathbb{L}}_{k})+\beta_{k}(\theta_{k}+e^{\mathbb{T}}_{k})-\theta_{k}\} \\
&=\alpha_{k}{\rm {E}}\{e^{\mathbb{L}}_{k}\}+\beta_{k}{\rm{E}}\{e^{\mathbb{T}}_{k}\}=\kappa\sigma_{T}^2\beta_{k}.
\end{split}
\end{equation}

Similarly, the correlation of $e^{\mathbb{F}}_{k}$ is further calculated as:
\begin{equation}
\begin{split}
&{\rm{E}}\{(e^{\mathbb{F}}_{k})^2\}={\rm{E}}\{(\alpha_{k}e^{\mathbb{L}}_{k}+\beta_{k}e^{\mathbb{T}}_{k})^2\}\\
&=\alpha_{k}^2{\rm{E}}\{(e^{\mathbb{L}}_{k})^2\}+\beta_{k}^2{\rm{E}}\{(e^{\mathbb{T}}_{k})^2\}+2\alpha_{k}\beta_{k}{\rm{E}}\{e^{\mathbb{L}}_{k}\cdot e^{\mathbb{T}}_{k}\}.  \\
\end{split}
\end{equation}

Since the causes of errors from each synchronization phase are not correlated, $e^{\mathbb{L}}_{k}$ and $e^{\mathbb{T}}_{k}$ are independent of each other and satisfy ${\rm{E}}\{e^{\mathbb{L}}_{k}\cdot e^{\mathbb{T}}_{k}\}={\rm{E}}\{e^{\mathbb{L}}_{k}\}\cdot{\rm{E}}\{e^{\mathbb{T}}_{k}\}$. Given the correlation analysis in (18) and (20), the fusion estimation error correlation in (23) is calculated as:
\begin{equation}
\begin{split}
{\rm{E}}\{(e^{\mathbb{F}}_{k})^2\}=\epsilon_k\alpha_{k}^2+\kappa^2\Big(4\sigma_{T}^2(T_k-T_{0})^2+3\sigma_{T}^4\Big)\beta_{k}^2.
\end{split}
\end{equation}

Combined with the above analysis, the variance of the fusion-based estimation error is calculated as:
\begin{equation}
\begin{split}
&({\sigma}^{\mathbb{F}}_{k})^2={\rm{E}}\{(e^{\mathbb{F}}_{k})^2\}-({\rm{E}}\{e^{\mathbb{F}}_{k}\})^2\\
&=\epsilon_k\alpha_{k}^2+\kappa^2\Big(4\sigma_{T}^2(T_k-T_{0})^2+2\sigma_{T}^4\Big)\beta_{k}^2.
\end{split}
\end{equation}

The estimation bias and variance should be minimized simultaneously in selecting fusion weight values, to reduce and achieve a trade-off between the average level and dispersion degree of estimation error as much as possible. The limitations of each clock model can be compensated mutually through the fusion of skew estimates, and the synchronization performance can be improved. The weight values are chosen periodically based on Pareto optimization \cite{7102992}, and the cost function is:  
\begin{equation}
\begin{split}
&\mathop{\rm{min}} \ \lambda_{k}(\mu^{\mathbb{F}}_{k})^2+(1-\lambda_{k})(\sigma^{\mathbb{F}}_{k})^2, \\
&{\rm{s.t.}} \ \beta_{k} \in [0, 1], \ \lambda_{k} \in [0, 1],
\end{split}
\end{equation}
where $\lambda_{k}$ is the Pareto weighting factor in the \emph{$k$}-th period. 

Given the fusion-based estimation error analysis in (22) and (25), the optimization object in (26) is converted to a convex quadratic function, and the extreme point of $\beta_{k}$ with Pareto weighting factor $\lambda_{k}$ can be expressed as:
\begin{equation}\nonumber
\begin{split}
\eta_k=\frac{\!(1-\lambda_{k})\cdot \epsilon_k}{\lambda_{k}\kappa^2\sigma_{T}^4\!+\!(1\!-\!\lambda_{k})\Big(\!\kappa^2\big(4(T_k\!-\!T_{0})^2\sigma_{T}^2\!+\!2\sigma_{T}^4\big)\!+\!\epsilon_k\!\Big)\!}.\!
\end{split}
\end{equation}

Thus, the optimal solution of fusion weight value $\beta_{k}$ is:
\begin{equation}
\begin{split}
\beta^*_{k}={\rm{max}}\big(0,\ {\rm{min}}(\eta_k,\ 1)\big).
\end{split}
\end{equation}

The optimal solution of $\beta_{k}$ depends on the value of $\lambda_{k}$. Many studies focus on selecting the optimum Pareto weight factor \cite{7102992}. In practical applications, the value of $\lambda_{k}$ can be set according to various requirements, e.g., when $\lambda_{k}$ is 0.5, the optimization objective is to minimize the mean square error. According to the actual value of factor $\lambda_{k}$, the weight values $\alpha_{k}$ and $\beta_{k}$ for fusion can be obtained. 

To sum up, through the limitation analysis for the two synchronization phases, the defects of each clock model could be compensated mutually. The estimation results are fused based on the Pareto optimization to balance the estimation bias and various simultaneously. The data fusion phase can adapt to dynamic networks and work environments.

\section{Theoretical Optimal Accuracy Analysis}

For evaluating the model fusion effect, the performance comparison between synchronization based on a single clock model and TACD should be studied. Bayesian Cramér-Rao lower bound (BCLB) quantifies the bound of estimation mean square error and characterizes the best accuracy that can be achieved in theory. We use BCLB as the comparison standard. 

The optimal accuracy of linear clock model-based synchronization is first analyzed. The process of estimating clock skew in the network communication phase is modeled as:
\begin{equation}
\begin{split}
\theta_k&=m\theta_{k-1}+u_k,\\
z_k&=h\theta_k+r_k,
\end{split}
\end{equation}
where equations are from (4) and (8). Measurement $z_k=t_{k}^{\mathsf{2}}-t_{k-1}^{\mathsf{2}}-t_{k}^{\mathsf{1}}+t_{k-1}^{\mathsf{1}}$, and coefficient $h=\tau$. Noise $r_k=w_{k}^{\mathsf{1}}-w_{k-1}^{\mathsf{1}}$ is the PDV in the same direction. Accordingly, $r_k$ is also modeled as non-stationary and has the same number of components as $\mathbf{n}_k$. The distribution of $r_k$ is expressed as:
\begin{equation}
\begin{split}
p(r_k)=\sum\nolimits_{j=1}^{N_g}b^j_k\mathcal{N}\big(r_k|0, (\Lambda^j_k)^2\big),\ \sum\nolimits_{j=1}^{N_g}b^j_k=1,  
\end{split}
\end{equation}
where \emph{$b^j_{k}$} and \emph{$\Lambda^j_k$} are the weight and standard deviation of the \emph{j}-th Gaussian component in the \emph{$k$}-th period, respectively. 

The Fisher information matrix is the inverse of BCLB. Since asymptotic approximation of variational inference, assuming complete knowledge of $r_k$ is known, we give the proposition:
\newtheorem*{proposition}{Proposition 1}
\begin{proposition}
The iterative expression of Fisher information matrix $\mathbf{J}^{\mathbb{L}}_k$ for representing the optimal accuracy of the linear clock model-based skew estimation is:
\begin{equation}
\begin{split}
\mathbf{J}^{\mathbb{L}}_{k}=\frac{1}{\sigma_{u}^2}-\frac{m}{\sigma_{u}^2}(\mathbf{J}^{\mathbb{L}}_{k-1}+\frac{m^2}{\sigma_{u}^2})^{-1}\frac{m}{\sigma_{u}^2}+\frac{\sum\nolimits_{j=1}^{N_g}\frac{b^j_{k}\tau^2}{(\Lambda^j_{k})^3}}{\sum\nolimits_{j=1}^{N_g}\frac{b^j_{k}}{\Lambda^j_{k}}}.
\end{split}
\end{equation}
\end{proposition}
\emph{Proof.} Please see Appendix A.

The estimation accuracy (30) cannot be achieved without the prior knowledge of $r_k$, which is unavailable in reality. Thus, the characterized performance is an estimation lower bound.

The optimal accuracy of TACD is then analyzed, whose skew estimation process is modeled as non-linear:
\begin{equation}
\begin{split}
&\mathbf{p}_{k}=\mathbf{m}(\mathbf{p}_{k-1})+\mathbf{u}_k,\\
&\mathbf{q}_{k}=\mathbf{F}\mathbf{p}_{k}+\mathbf{r}_k,
\end{split}
\end{equation}
where $\mathbf{p}_k=\begin{bmatrix}  \   \theta_k    \ \  \\  \   T_k   \ \  \end{bmatrix}$, $\mathbf{u}_k=\begin{bmatrix}  \   \alpha_{k}u_k    \ \  \\  \    v_k   \ \  \end{bmatrix}$. \emph{$v_k$} is the temperature increment during operation, modeled as Gaussian with mean 0 and variance \emph{$\sigma_{m}^2$}. Thus $\mathbf{u}_k$ is a Gaussian noise vector with mean vector $\bm{0}$ and covariance matrix \emph{$\mathbf{Q}_{u}=\begin{bmatrix}  \   \alpha^2_{k}\sigma_{u}^2  &  0  \ \  \\  \   0  &  \sigma_{m}^2 \ \  \end{bmatrix}$}. The non-linear state update function is: 
\begin{equation}\nonumber
\begin{split}
\!\mathbf{m}(\mathbf{p}_{k})\!=\!\begin{bmatrix}  \   \!\alpha_k\cdot m\theta_{k}+\beta_k\cdot \big(\kappa(T_k\!-\!T_{0})^2\!+\!\theta_{0}\big)\!  \ \  \\  \   T_k \!\ \  \end{bmatrix}\!
\end{split}
\end{equation}

In addition, $\emph{$\mathbf{F}=\begin{bmatrix}  \   \tau  &  0  \ \  \\  \   0  &  1 \ \  \end{bmatrix}$}$, $\emph{$\mathbf{r}_k=\begin{bmatrix}  \  \! \alpha_k(w_{k}^{\mathsf{1}}\!-\!w_{k-1}^{\mathsf{1}}) \! \ \  \\  \   \xi_{k} \ \  \end{bmatrix}$}$. Since the non-stationary noise $r_k$ is included in \emph{$\mathbf{r}_k$}, the probability distribution of the noise matrix \emph{$\mathbf{r}_k$} is:
\begin{equation}
\begin{split}
p(\mathbf{r}_k)=\sum\nolimits_{j=1}^{N_g}b^j_k\mathcal{N}(\mathbf{r}_k|\mathbf{0}, \mathbf{\Upsilon}^j_k),\ \sum\nolimits_{j=1}^{N_g}b^j_k=1,  \\
\end{split}
\end{equation}
where $\mathbf{\Upsilon}^j_k$ is the covariance matrix of the \emph{$j$}-th Gaussian component. Considering there is no correlation between terms $\alpha_kr_k$ and $\xi_{k}$ in noise \emph{$\mathbf{r}_k$}, so $\emph{$\mathbf{\Upsilon}^j_k=\begin{bmatrix}  \alpha^2_k(\Lambda^j_k)^2  &  0    \\    0  &  \sigma_{T}^2   \end{bmatrix}$}$. The measurement $\mathbf{q}_{k}$ combining the two phases is expressed as:
\begin{equation}\nonumber
\begin{split}
\begin{bmatrix} \!\alpha_k(t_{k}^{\mathsf{2}}\!-\!t_{k-1}^{\mathsf{2}}\!-\!t_{k}^{\mathsf{1}}\!+\!t_{k-1}^{\mathsf{1}})\!+\!\beta_k\tau\big(\kappa(\tilde{T}_k\!-\!T_{0})^2\!+\!\theta_{0}\big) \! \\     \tilde{T}_k    \end{bmatrix}
\end{split}
\end{equation}

Assuming the complete knowledge of $\mathbf{r}_k$ is known, we give:

\newtheorem*{proposition2}{Proposition 2}
\begin{proposition2}
The iterative expression of Fisher information matrix $\mathbf{J}^{\mathbb{F}}_k$ for representing the optimal accuracy of the fusion-based skew estimation is: 
\begin{equation}\nonumber
\begin{split}
\!\mathbf{J}^{\mathbb{F}}_{k}\!=\!\frac{1}{\!\alpha^2_{k-1}\sigma_{u}^2}\!-\!\frac{m}{\!\alpha_{k-1}\sigma_{u}^2}(\mathbf{J}^{\mathbb{F}}_{k-1}\!+\!\frac{m^2}{\sigma_{u}^2})^{\!-1}\!\frac{m}{\!\alpha_{k-1}\sigma_{u}^2}\!+\!\frac{\!\sum_{j=1}^{N_g}\frac{b^j_{k}\tau^2}{(\Lambda^j_{k})^3}}{\!\alpha^2_{k}\!\sum_{j=1}^{N_g}\frac{b^j_{k}}{\Lambda^j_{k}}}.
\end{split}
\end{equation}
\end{proposition2}
\emph{Proof.} Please see Appendix B.

From this, the theoretical optimal synchronization accuracy based on a single clock model and model fusion is analyzed, laying a foundation for further evaluation. Through comparison, the weight value $\alpha_{k}$ is introduced into the Fisher information matrix after model fusion. Considering $\alpha_{k} \in [0, 1]$, it can be intuitively concluded that model fusion reduces BCLB, thereby improving optimal accuracy. 

\begin{figure} [t]
  \centering
    \subfigure[\label{fig:a}]
{\includegraphics[width=5.6cm]{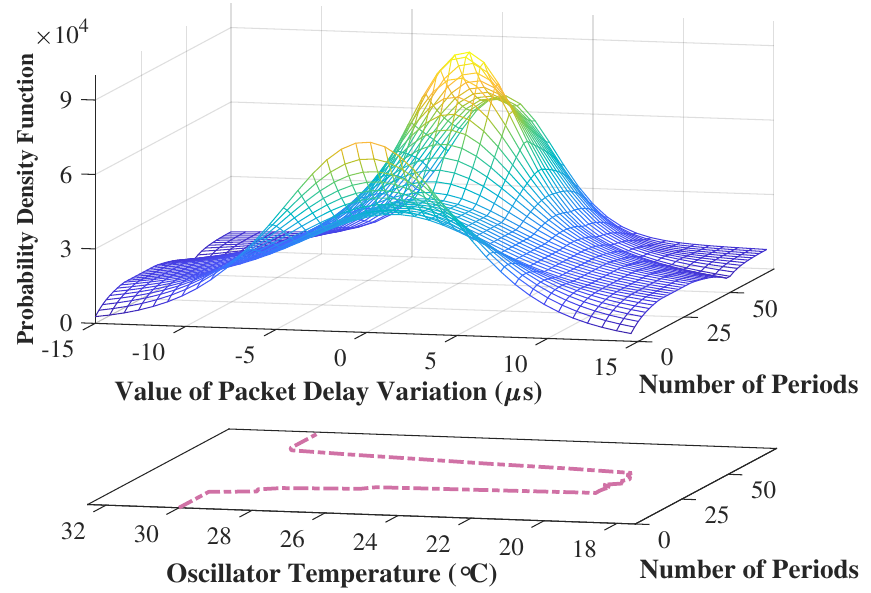}} 
\centering
    \subfigure[\label{fig:b}]
{\includegraphics[width=5.35cm]{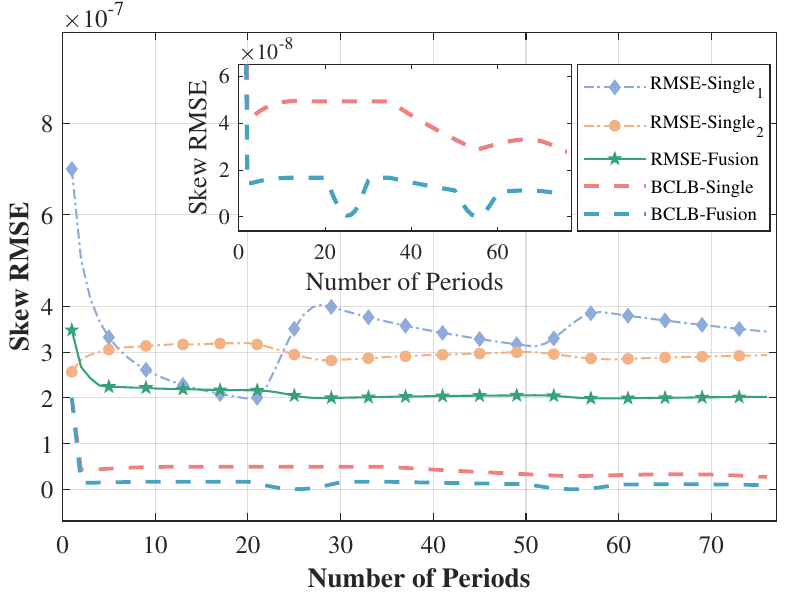}}  
\vspace{-0.15cm}
  \caption{Fusion performance investigation. (a) Changing trends of delay distributions and temperature values in the simulation; (b) Comparison of effects based on various clock models.}
	\label{fig:data_distribution}
\vspace{-0.4cm}
\end{figure}

\begin{figure} [t]
  \centering
    \subfigure[\label{fig:a}]
{\includegraphics[width=5.1cm]{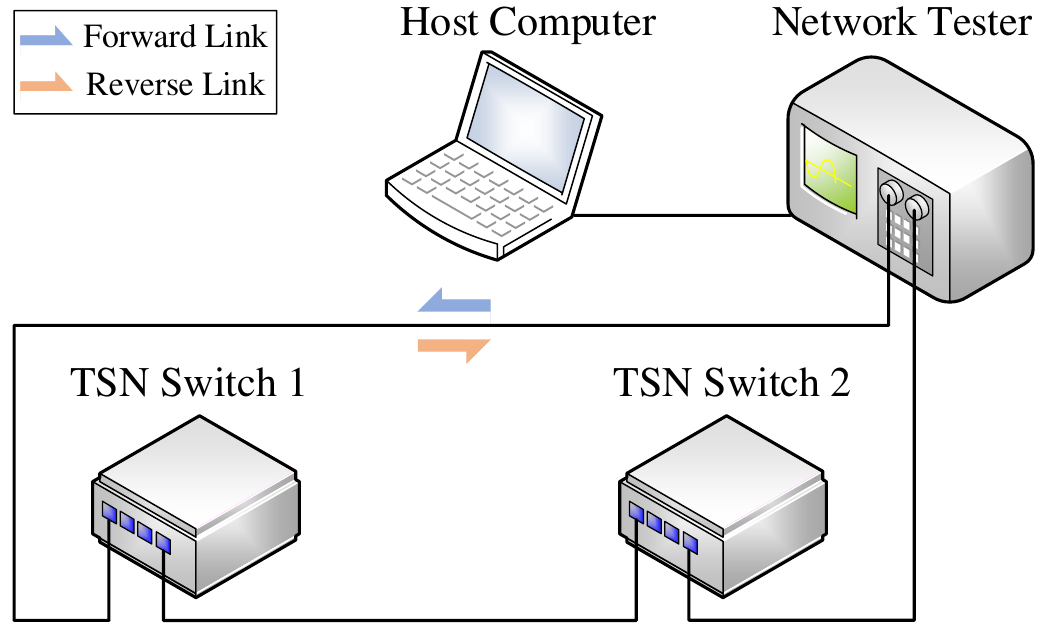}} 
\centering
    \subfigure[\label{fig:b}]
{\includegraphics[height=3.1cm]{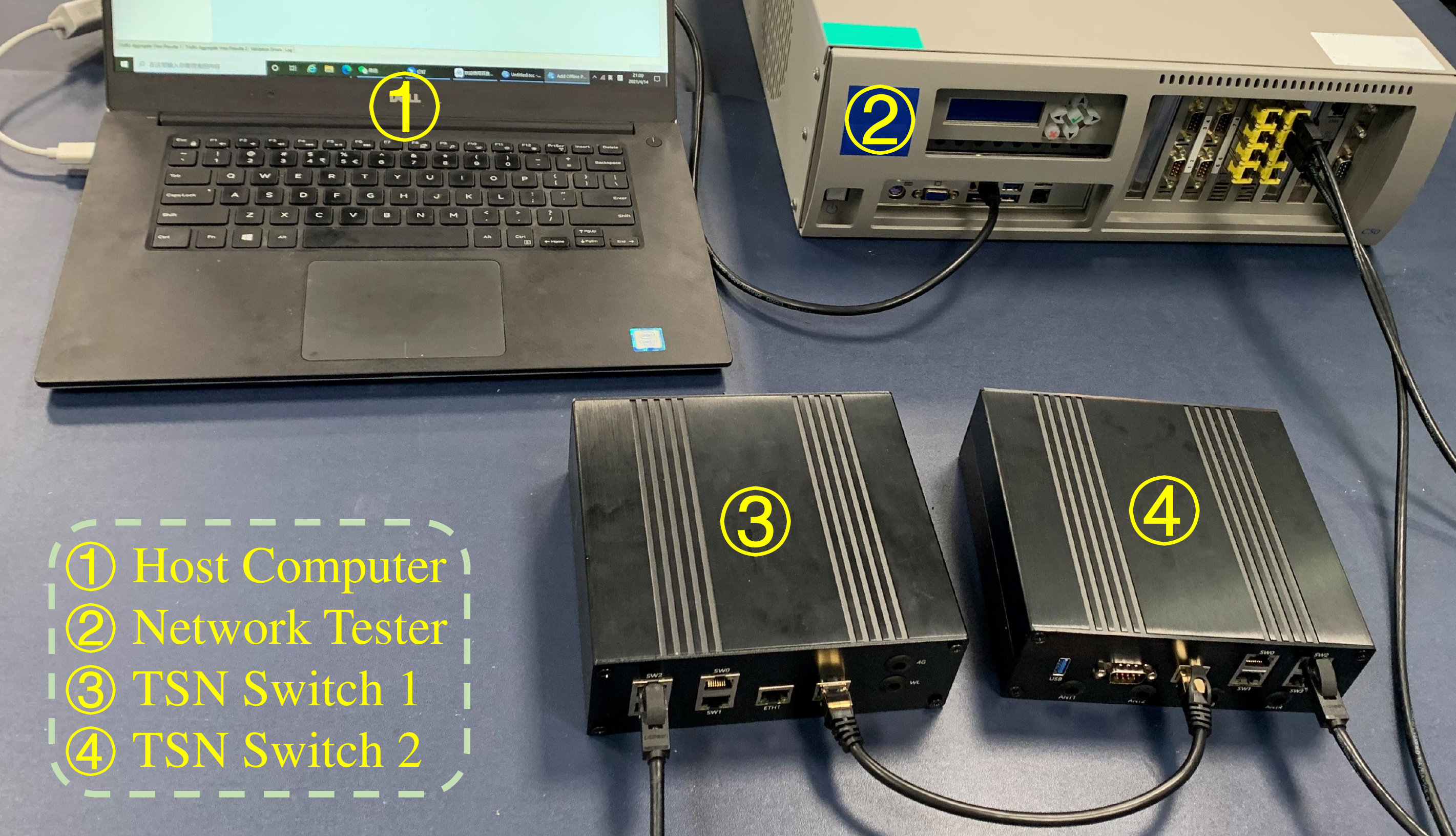}}  
\centering
    \subfigure[\label{fig:c}]
{\includegraphics[height=4.5cm]{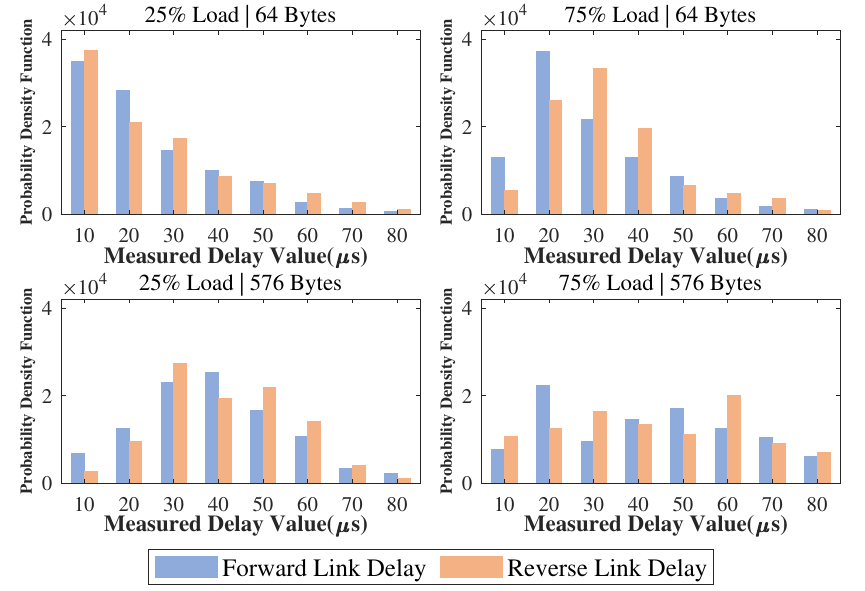}}
\vspace{-0.15cm}
  \caption{Delay data acquisition. (a) The topology for obtaining delay data; (b) The scenario of collecting delay data; (c) Partial delay distributions sampled under different network status.}
	\label{fig:data_distribution}
\vspace{-0.45cm}
\end{figure}

\section{Performance Evaluation}

This section studies the performance of TACD. Specifically, the model fusion effect is first evaluated compared to using a single clock model. The synchronization effect of TACD is then compared with other methods based on actual measured delay data. The resource consumption is also analyzed.

\begin{table}[b]
\centering
\scriptsize
\caption{Parameter Settings for the Study Case}
\label{tab:table1}
\renewcommand{\arraystretch}{1.3}
\begin{tabular}{m{0.9cm}|m{1.03cm}<{\centering} m{1.03cm}<{\centering} m{1.03cm}<{\centering} m{1.03cm}<{\centering} m{1.03cm}<{\centering}c c c c c c}
\toprule[0.45pt]
\multirow{2}*{\textbf{\makecell{Initial \\ Values}} }  &  $\scriptsize{{\Lambda^1_0}}_{\tiny{(\!\times\!10^{-6})}}$  &  $\scriptsize{{\Lambda^2_0}}_{\tiny{(\!\times\!10^{-6})}}$  &  $\scriptsize{{\Lambda^3_0}}_{\tiny{(\!\times\!10^{-6})}}$  &  $\scriptsize{{b^1_0}}$  &  $\scriptsize{{b^2_0}}$  \\ 
\cline{2-6}
~  &  $\text{5}$  &  $\text{3}$  &  $\text{5}$  &  $\text{0.4}$  &  $\text{0.3}$  \\
\end{tabular}
\renewcommand{\arraystretch}{0.2}
\begin{tabular}{m{0.9cm}|m{1.03cm}<{\centering} m{1.03cm}<{\centering} m{1.03cm}<{\centering} m{1.03cm}<{\centering} m{1.03cm}<{\centering}c c c c}
\toprule[0.45pt]
\textbf{Changing Rates}  &  $\scriptsize{{n^1_k}}_{\tiny{(\!\times\!10^{-7})}}$  &  $\scriptsize{{n^2_k}}_{\tiny{(\!\times\!10^{-7})}}$  &  $\scriptsize{{n^3_k}}_{\tiny{(\!\times\!10^{-7})}}$  &  $\scriptsize{{n^4_k}}_{\tiny{(\!\times\!10^{-3})}}$  &  $\scriptsize{{n^5_k}}_{\tiny{(\!\times\!10^{-3})}}$ \\ 
\midrule[0.45pt]
$_{  \tiny{1\leqslant k\leqslant15}}$  &  $\text{-0.6}$  &  $\text{2.4}$ & $\text{8}$  &  $\text{-11.6}$  &  $\text{-2.8}$ \\
$_{\tiny{16\leqslant k\leqslant35}}$  &  $\text{0}$  &  $\text{0}$	 &  $\text{0}$	 &  $\text{0}$  &  $\text{0}$\\  
$_{\tiny{36\leqslant k\leqslant55}}$  &  $\text{-7.25}$  &  $\text{-1.1}$  &  $\text{-5.5}$  &  $\text{11.7}$  &  $\text{-4}$ \\
$_{\tiny{56\leqslant k\leqslant75}}$  &  $\text{4.5}$  &  $\text{-1.3}$	 &  $\text{7}$ 	&  $\text{-11.5}$	 &  $\text{-3.5}$ \\ 
\bottomrule[0.45pt] 
\end{tabular}
\renewcommand{\arraystretch}{0.2}
\begin{tabular}{m{1.25cm}|m{1.85cm}<{\centering} m{4.3cm}<{\centering} c c c}
\toprule[0.0pt]
\textbf{Temperature Variation}  &  \textbf{Curve Type}  &  \textbf{Model Expression}   \\ 
\midrule[0.45pt]
$_{\tiny{k\leqslant20 \ \& \ 61\leqslant k}}$ &  \textrm{Constant Value}  &  $\tiny{30}$   \\
$_{\tiny{21\leqslant k\leqslant30}}$ &\textrm{Multimodal Curve}&$\tiny{1.1{\rm{sin}}(2k\!+\!\pi)\!-\!0.005(2k\!+\!2)^2\!+\!40}$\\  
$_{\tiny{31\leqslant k\leqslant50}}$  &  \textrm{Colored Noise}  &  $\tiny{20\!+\!\mathcal{N}\big(0,\! \  0.02\!+\!(k\!-\!30)\!\times\!10^{-2}\big)}$   \\
$_{\tiny{51\leqslant k\leqslant60}}$  &  \textrm{First-Order Curve}  &  $\tiny{k-30}$	\\  
\bottomrule[0.45pt]
\end{tabular}
\end{table}

\subsection{Fusion Performance Investigation}

Since the derived optimal accuracy depends on delay distribution parameters not available in practice, the performance of model fusion is investigated through numerical simulations. 

\subsubsection{Simulation Setup}

To simulate dynamic networks and environments in industrial TSN, we consider a case where delay variation and temperature values are non-stationary time-varying, so the effect under internal and external changes is compared. Parameters of PDV model $r_k$ in (29) are set as:
\begin{equation}
\begin{split}
&N_g=3; \ \Lambda^j_k=\Lambda^j_0+n^j_k\cdot k, \ j=1, 2, 3, \\
b^1_k=&b^1_0+n^4_k\cdot k, \ b^2_k=b^2_0+n^5_k\cdot k, \ b^3_k=1-b^1_k-b^2_k,
\end{split}
\end{equation}
where $n^j_k$ is the parameter changing rate. Parameter settings of delay variation and oscillator temperature variation are shown in Table \uppercase\expandafter{\romannumeral1}, and changing trends are illustrated in Fig. 2(a).

Other parameters include $m=(1-2\times10^{-6})^{\frac{1}{30}}$ and $\sigma_u^2=(1-m^2)\times10^{-3}$ according to \cite{9475453}. Period $\tau$ is $1$s as PTP default settings \cite{9013260}. Consistent with \cite{9081999}, $d_1$ and $d_2$ are 5 and 1${\mu}$s, and the initial offset $\delta_0$ is 1${\mu}$s. We set $\kappa=0.04$ppm and $T_{0}=25^{\circ}$C in line with the clock model in \cite{6836139, 8125129}. The value of $\sigma_{T}^2$ is set as 0.1 \cite{6836139}. Variational inference is initialized with ${\bm{\chi}}_{0}=[1, 5, 5]$, ${\upsilon}_{0}=[4, 3, 3]$, and ${\mathbf{V}}_{0}=\rm{diag}(10^{5}, 2\times10^{5}, 2\times10^{5})$. Initialize the estimation with ${\mathbf{x}}_{0}=[3\times10^{-7},  \ 3.5\times10^{-6}]^{\rm{T}}$ and ${\mathbf{P}}_{0}=\rm{diag}(5\times10^{-6}, 5\times10^{-6})$. The simulation was iterated for 75 periods and averaged for 1000 runs. 

\begin{figure*} [t]
  \centering
{\begin{minipage}[t]{0.3\textwidth}
    \subfigure[\label{fig:a}]
{\includegraphics[width=5.4cm]{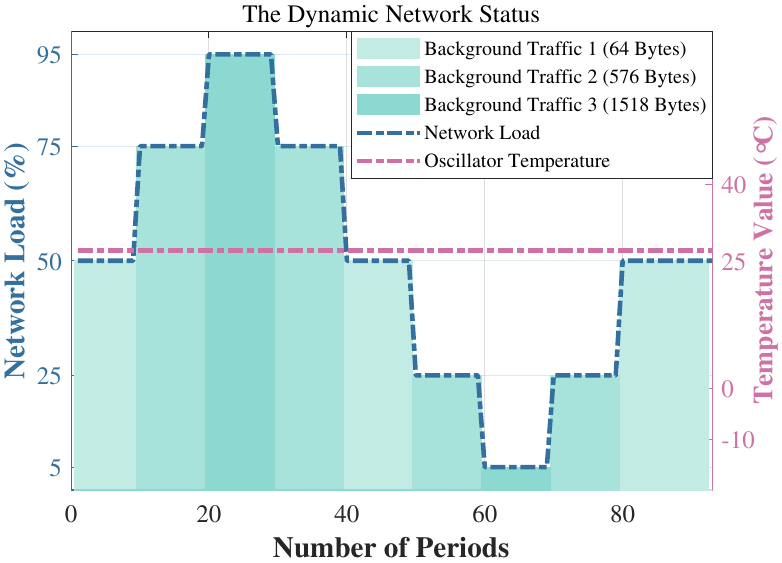}} 
\end{minipage}}
\centering
{\begin{minipage}[t]{0.3\textwidth}
    \subfigure[\label{fig:b}]
{\includegraphics[width=5.1cm]{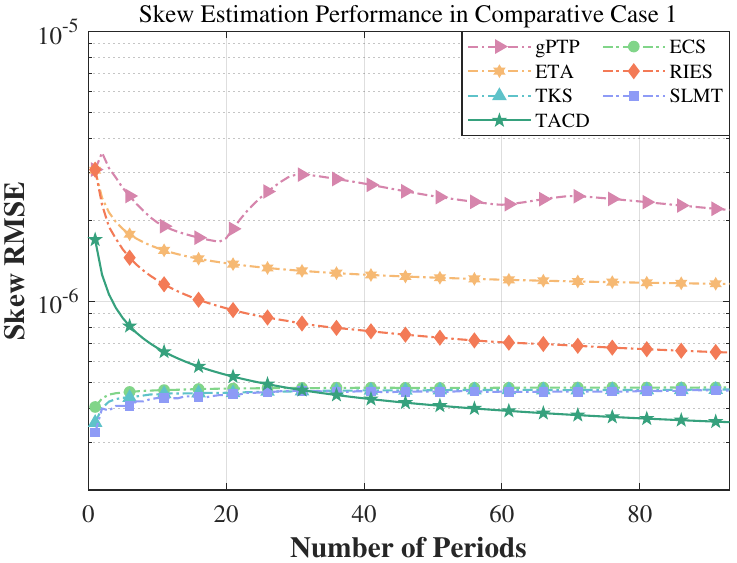}}  
\end{minipage}}
{\begin{minipage}[t]{0.3\textwidth}
    \subfigure[\label{fig:b}]
{\includegraphics[width=5.1cm]{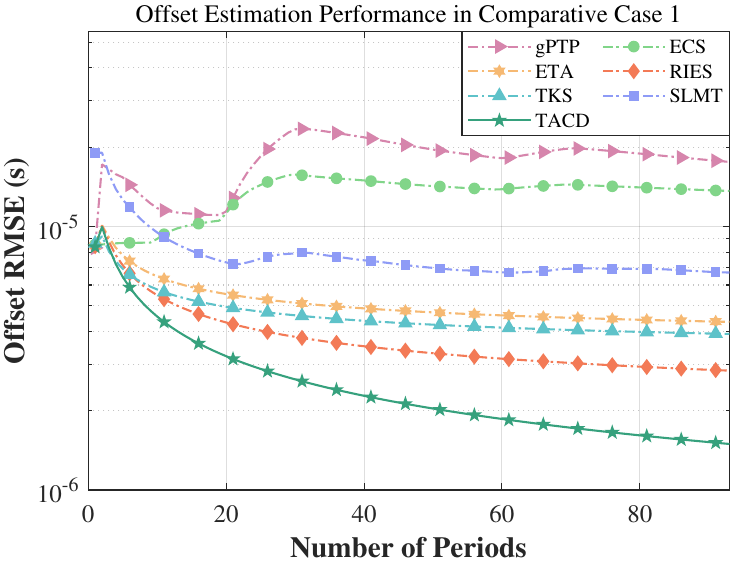}}  
\end{minipage}}
\vspace{-0.2cm}
  \caption{Comparison of synchronization performance in comparative case 1. (a) The simulated network status and ambient temperature; (b) Comparison of skew estimation effects; (c) Comparison of offset estimation effects.}
	\label{fig:data_distribution}
\end{figure*}
\begin{figure*} [t]
  \centering
{\begin{minipage}[t]{0.3\textwidth}
    \subfigure[\label{fig:a}]
{\includegraphics[width=5.4cm]{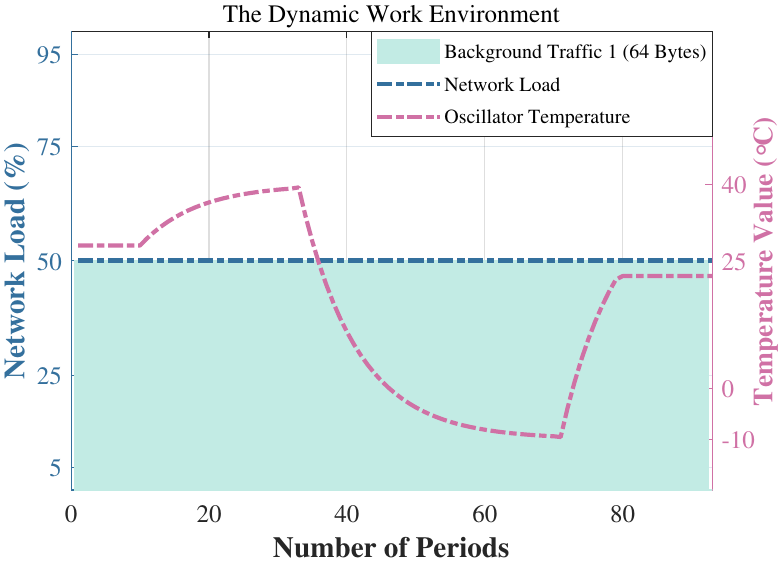}} 
\end{minipage}}
\centering
{\begin{minipage}[t]{0.3\textwidth}
    \subfigure[\label{fig:b}]
{\includegraphics[width=5.1cm]{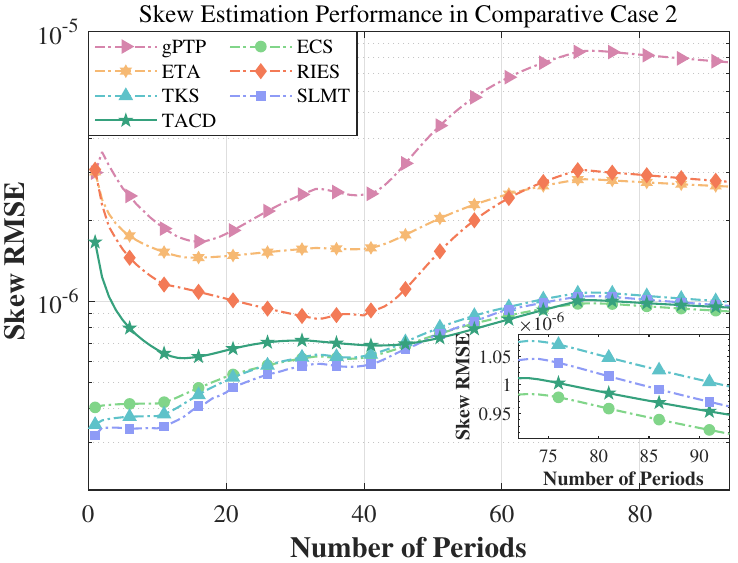}}  
\end{minipage}}
{\begin{minipage}[t]{0.3\textwidth}
    \subfigure[\label{fig:b}]
{\includegraphics[width=5.1cm]{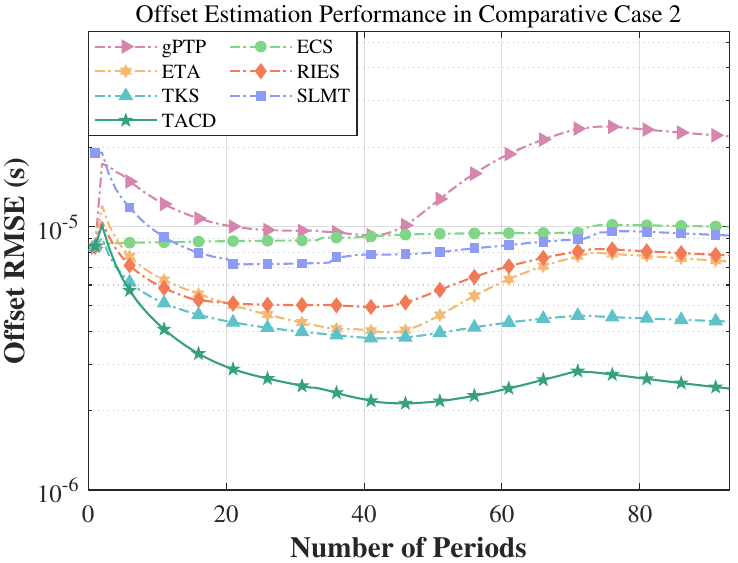}}  
\end{minipage}}
\vspace{-0.2cm}
  \caption{Comparison of synchronization performance in comparative case 2. (a) The simulated network status and ambient temperature; (b) Comparison of skew estimation effects; (c) Comparison of offset estimation effects.}
\vspace{-0.3cm}
	\label{fig:data_distribution}
\end{figure*}

\subsubsection{Simulation Results}

Fig. 2(b) depicts root mean squared error (RMSE) and BCLB performance of skew estimation based on various models. For estimation based on a single linear clock model (``RMSE-Single$_{1}$''), nonlinear errors from temperature changes seriously reduce its effect. Synchronization based on a single temperature skew model (``RMSE-Single$_{2}$'') adapts to thermal changes but is limited by measurement accuracy. Fusion-based estimator (``RMSE-Fusion'') simultaneously overcomes each model's limitations and performs better. From steady-state results in the last ten periods, the theoretical optimal accuracy of the proposed skew estimation (``BCLB-Fusion'') is 66\% lower than that based on the single linear clock model (``BCLB-Single''). Thus, the fusion of clock models brings better skew estimation effects.
\subsection{Performance Comparison with Existing Methods}

\subsubsection{Data in Simulation}

Topology to obtain delay data under various network statuses is built, shown in Fig. 3(a). The host computer used to issue sampling instructions contains an Intel(R) i7-7700HQ CPU with 8 GB of RAM. The network tester used to transmit and receive traffic flows and measure propagation delay is Spirent C50. Network nodes are switches with LS1028A chips and support TSN mechanisms. Fig. 3(b) shows the scenario of collecting delay data.

In delay measurement, traffic flows are injected into the link through the tester output port, comprising lower-priority background flows and higher-priority synchronization messages. Various network statuses are simulated by setting different background traffic packet sizes and loads. The traffic packet sizes specified in ITU-T G.8261 standard for testing are \{64, 576, 1518\} bytes \cite{8643987}. The tester can set the load \{5, 25, 50, 75, 95\}\%. The measured delay data is for the following simulations, whose partial distributions are shown in Fig. 3(c).

To simulate the oscillator temperature variation, we select $-10^{\circ}$C to $40^{\circ}$C as the variation range according to \cite{5981981, 6817598}, meeting the thermal test requirements of switches. The actual variation trend satisfies Newton's law of cooling \cite{6817598}:
\begin{equation}
\begin{split}
T_{{\rm{o}}}(t)=T_{{\rm{e}}}(t)+\big(T_{{\rm{o}}}(0)-T_{{\rm{e}}}(t)\big)\cdot e^{-\frac{t}{c_{\rm{t}}}},
\end{split}
\end{equation}
where $T_{{\rm{o}}}(0)$ is the initial oscillator temperature, and $T_{{\rm{e}}}(t)$ is the changing external temperature. The thermal time constant $c_{\rm{t}}$ is 10, and the thermal period is set to 75s.

\begin{figure*} [t]
  \centering
{\begin{minipage}[t]{0.3\textwidth}
    \subfigure[\label{fig:a}]
{\includegraphics[width=5.4cm]{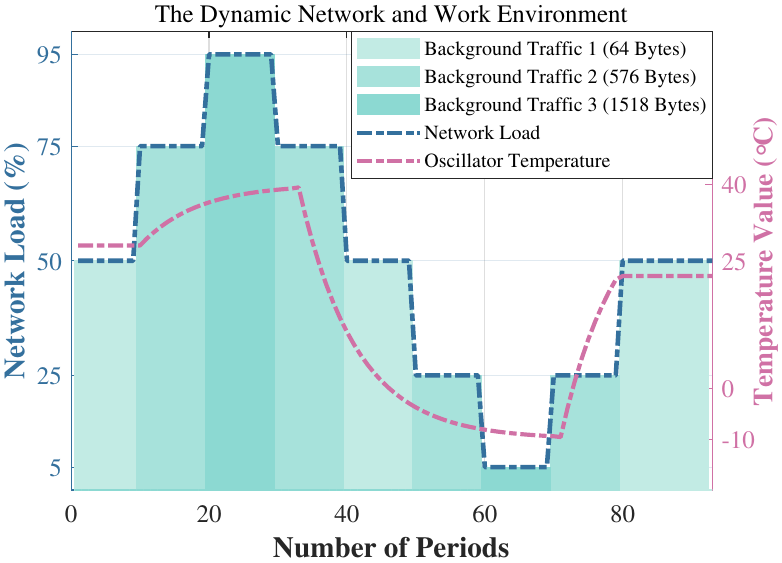}} 
\end{minipage}}
\centering
{\begin{minipage}[t]{0.3\textwidth}
    \subfigure[\label{fig:b}]
{\includegraphics[width=5.1cm]{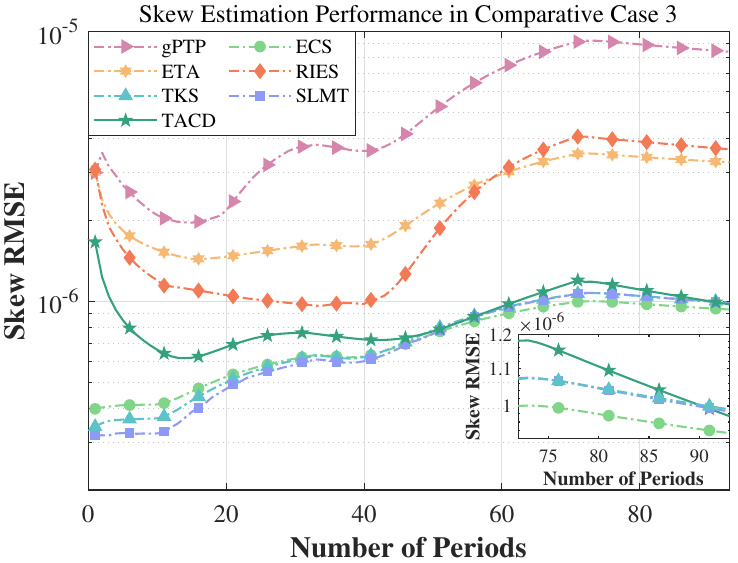}}  
\end{minipage}}
{\begin{minipage}[t]{0.3\textwidth}
    \subfigure[\label{fig:b}]
{\includegraphics[width=5.1cm]{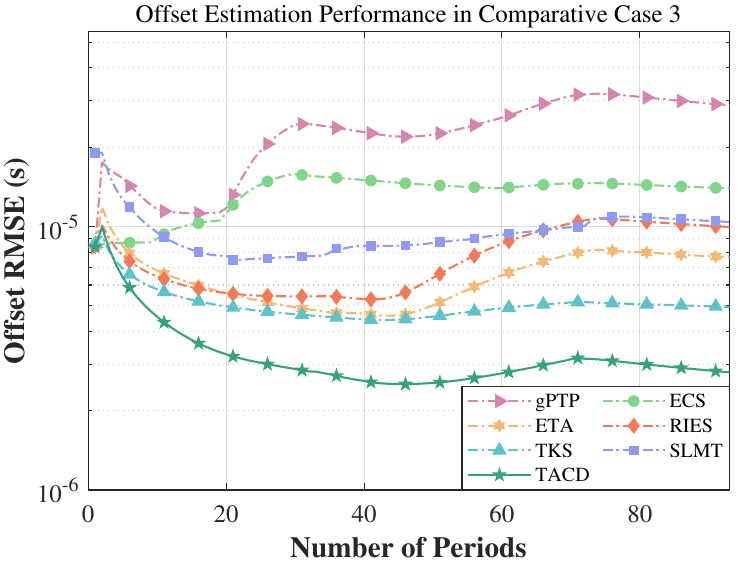}}  
\end{minipage}}
\vspace{-0.2cm}
  \caption{Comparison of synchronization performance in comparative case 3. (a) The simulated network status and ambient temperature; (b) Comparison of skew estimation effects; (c) Comparison of offset estimation effects.}
\vspace{-0.3cm}
	\label{fig:data_distribution}
\end{figure*}

\subsubsection{Methods for Comparison}

The performance of TACD is compared with that of gPTP (method of 802.1AS in TSN) \cite{9121845}, RIES (robust iterative estimation scheme) \cite{9081999}, ETA (extended Kalman filter-based tracking algorithm) \cite{9475453}, SLMT (synchronization using linear programming, multicasts, and temperature compensation) \cite{9013260}, ECS (external clock synchronization method) \cite{6836139} and TKS (temperature-compensated Kalman-based synchronization) \cite{GONG201788}. Among them, gPTP, RIES, and ETA are based on two-way communication, and SLMT, ECS, and TKS consider the clock's operation characteristics.

\subsubsection{Comparison in Dynamic Networks}

Comparative case 1 studies the network impact. Specifically, a dynamic network is simulated by setting time-varying loads and packet sizes of background traffic. Temperature is set to be fixed at $28^{\circ}$C for variable control. Fig. 4(a) shows parameter settings of network status and oscillator temperature. Fig. 4(b) and (c) show each method's skew and offset estimation performance. Simulations were iterated for 95 periods and averaged for 1000 runs. 

For skew estimation, the proposed TACD performs best after prior iterations. Table \uppercase\expandafter{\romannumeral2} shows steady-state average RMSEs over the last ten periods. The RMSE of TACD is 22\% lower than that of SLMT, which acts best among existing methods. For offset estimation, methods modeling PDVs with probability distributions are more adapted to dynamic networks. Among them, RIES using hybrid models has better accuracy. Compared with RIES, TACD is reduced by 46\% in RMSE.

\begin{table}[b]
\centering
\scriptsize
\vspace{-0.2cm}
\caption{Steady-State RMSEs of Various Methods}
\label{tab:table1}
\renewcommand{\arraystretch}{1.4}
\begin{threeparttable}
\begin{tabular}{m{0.90cm}|m{0.81cm}<{\centering} m{0.81cm}<{\centering} m{0.81cm}<{\centering}| m{0.81cm}<{\centering} m{0.81cm}<{\centering} m{0.81cm}<{\centering} c c c c c c}
\toprule[0.45pt]
\multirow{2}*{\textbf{\makecell{Methods}} }  & \multicolumn{3}{c|}{ ${\textbf{\makecell{Skew}}(\!\times\!10^{-7})}$}   & \multicolumn{3}{c}{ ${\textbf{\makecell{Offset}}(\!\times\!10^{-6})}$}    \\ 
\cline{2-7}
~  &  $\textrm{\makecell{Case 1}}$  &  $\textrm{\makecell{Case 2}}$  &  $\textrm{\makecell{Case 3}}$  &  $\textrm{\makecell{Case 1}}$  &  $\textrm{\makecell{Case 2}}$   &  $\textrm{\makecell{Case 3}}$  \\
\end{tabular}
\renewcommand{\arraystretch}{0.6}
\begin{tabular}{m{0.9cm}|m{0.81cm}<{\centering} m{0.81cm}<{\centering} m{0.81cm}<{\centering}| m{0.81cm}<{\centering} m{0.81cm}<{\centering} m{0.81cm}<{\centering} c c c c c c}
\toprule[0.45pt]
\textbf{gPTP}  &  $\text{22.31}$  &  $\text{78.62}$ & $\text{85.92}$  &  $\text{18.03}$  &  $\text{22.58}$  &  $\text{29.64}$ \\
\textbf{ETA}  &  $\text{11.62}$  &  $\text{27.03}$	 &  $\tiny{33.36}$	 &  $\text{4.37}$  &  $\text{7.50}$ &  $\text{7.76}$\\  
\textbf{RIES}  &  $\text{6.53}$  &  $\text{28.34}$  &  $\text{37.42}$  &  $\text{2.87}$  &  $\text{7.87}$ &  $\text{10.14}$\\
\textbf{TKS}  &  $\text{4.71}$  &  $\text{10.13}$	 &  $\text{10.17}$ 	&  $\text{3.93}$	 &  $\text{4.41}$ &  $\text{5.00}$\\ 
\textbf{SLMT}  &  $\text{4.67}$  &  $\text{9.81}$	 &  $\text{10.15}$ 	&  $\text{6.77}$	 &  $\text{9.36}$ &  $\text{10.64}$\\ 
\textbf{ECS}  &  $\text{4.79}$  &  $\textbf{9.32}$	 &  $\textbf{9.48}$ 	&  $\text{13.85}$	 &  $\text{10.03}$ &  $\text{14.12}$\\ 
\textbf{TACD}  &  $\textbf{3.61}$  &  $\text{9.61}$	 &  $\text{10.21}$ 	&  $\textbf{1.53}$	 &  $\textbf{2.50}$ &  $\textbf{2.87}$\\ 
\bottomrule[0.45pt] 
\end{tabular}
    \end{threeparttable}
  \label{tb:1}%
\renewcommand{\arraystretch}{0.7}
\end{table}

\subsubsection{Comparison in Dynamic Environments}

Comparative case 2 studies external environment impacts. The temperature range is set according to thermal tests in \cite{5981981, 6817598}, and oscillator temperature change meets the cooling law (34). Network status variation is not considered. Fig. 5(a) shows parameter settings. Fig. 5(b) and (c) show estimation performance.

In skew estimation, methods considering the clock's characteristics have lower RMSEs than those based on the linear clock model. Table \uppercase\expandafter{\romannumeral2} shows that TACD accuracy is quite close to those based on the temperature skew model, even when the temperature changes rapidly. In offset estimation, methods ignoring clock characteristics are also more severely affected by temperature. The RMSE of TACD is reduced by 43\%, compared with TKS performing best in existing methods.

\subsubsection{Comparison in Dynamic Networks and Environments}

The dual impact of internal and external is studied in case 3. Parameters are set by combining the prior two cases, as in Fig. 6(a). Fig. 6(b) and (c) depict the estimation results.

Compared with Fig. 5(b), non-stationary delay asymmetry seriously declines the effect of message exchange-based skew estimation, and temperature changes have greater impacts than dynamic networks. Table \uppercase\expandafter{\romannumeral2} shows that the skew estimation effect of TACD is quite close to ECS's (best in existing methods). In offset estimation, TKS acts optimally among existing methods, considering both clock's characteristics and PDV, and benefits from superior skew estimation. TACD acts better, with RMSE reduced by 42\% compared with TKS. 

\subsection{Resource Consumption Analysis}

For communication costs, TACD is based on TSN protocol, which utilizes existing network datagrams without extra traffic packets, so TACD generates no additional burden. For computation cost, TACD algorithm complexity is measured in terms of multiplications and is $\mathcal{O}(N_{g}^2)$, where $\mathcal{O}(.)$ is big-O notation. Computation cost is mainly from variational inference in the network communication phase. The increase of components number in model $\mathbf{n}_k$ leads to a complexity rise but also improves estimation accuracy. Variable $N_{g}$ can be set according to practical accuracy needs. Thus, TACD is feasible. 

\section{Conclusion}

This paper proposes a TACD architecture to promote synchronization performance in industrial TSN. TACD contains three phases: network communication phase, environment awareness phase, and data fusion phase. The partial variational Bayesian algorithm enables the two-way communication process to adapt to the impact of non-stationary delay asymmetry. With limitation analysis for each phase, an optimized skew estimator is designed to compensate for defects of each clock model. The simulation results show that the promotion of optimal accuracy could reach 66\% after the fusion of clock models in light of BCLB theoretical derivation. Based on measured delay data, TACD outperforms existing methods in three comparative cases. The feasibility is also analyzed. Future work will devote to synchronization reliability.

\section* {Appendix A \\ Proof of Proposition 1}

Fisher information matrix is recursively calculated as: 
\begin{equation}\nonumber
\begin{split}
\mathbf{J}_{k}=\mathbf{D}^{22}_{k-1}-\mathbf{D}^{21}_{k-1}(\mathbf{J}_{k-1}+\mathbf{D}^{11}_{k-1})^{-1}\mathbf{D}^{12}_{k-1},
\end{split}
\end{equation}
where $\mathbf{D}^{11}_{k-1}$, $\mathbf{D}^{12}_{k-1}$, and $\mathbf{D}^{22}_{k-1}$ are the Fisher components.

For skew estimation process (28), the conditional probability density function of state transition is:
\begin{equation}\nonumber
\begin{split}
p(\theta_{k}|\theta_{k-1})=\frac{1}{\sqrt{2\pi\sigma_{u}^2}}e^{-\frac{(\theta_{k}-m\theta_{k-1})^2}{2\sigma_{u}^2}},
\end{split}
\end{equation}
and the measurement likelihood distribution is also determined, whose conditional probability density function is: 
\begin{equation}\nonumber
\begin{split}
p(z_{k}|\theta_{k})=\sum_{j=1}^{N_g}b^j_{k}(\frac{1}{\sqrt{2\pi}\Lambda^j_{k}}e^{-\frac{(z_{k}-h\theta_{k})^2}{2(\Lambda^j_{k})^2}}).
\end{split}
\end{equation}

From there, the Fisher components can be calculated as:
\begin{equation}\nonumber
\begin{split}
\mathbf{D}^{11}_{k-1}=&-{\rm{E}}\{\nabla_{\theta_{k-1}}{\big(\nabla_{\theta_{k-1}}{\rm{log}}\ p(\theta_{k}|\theta_{k-1})\big)}^T\}
=\frac{m^2}{\sigma_{u}^2}, \\
\mathbf{D}^{12}_{k-1}=&-{\rm{E}}\{\nabla_{\theta_{k-1}}{\big(\nabla_{\theta_{k}}{\rm{log}}\ p(\theta_{k}|\theta_{k-1})\big)}^T\}
=-\frac{m}{\sigma_{u}^2},\\
\mathbf{D}^{22}_{k-1}=&-{\rm{E}}\{\nabla_{\theta_{k}}{\big(\nabla_{\theta_{k}}{\rm{log}}\ p(\theta_{k}|\theta_{k-1})\big)}^T\}\\
&-{\rm{E}}\{\nabla_{\theta_{k}}{\big(\nabla_{\theta_{k}}{\rm{log}}\ p(z_{k}|\theta_{k})\big)}^T\}\\
=&\frac{1}{\sigma_{u}^2}+\frac{\sum\nolimits_{j=1}^{N_g}\frac{b^j_{k}}{(\Lambda^j_{k})^3}h^2}{\sum\nolimits_{j=1}^{N_g}\frac{b^j_{k}}{\Lambda^j_{k}}}.
\end{split}
\end{equation}

Thus, the Fisher information matrix $\mathbf{J}^{\mathbb{L}}_{k}$ of clock skew estimation based on the linear clock model is obtained. 

\section* {Appendix B \\ Proof of Proposition 2}

For the state equation with Gaussian noise in the non-linear model (31), Fisher components are calculated as:
\begin{equation}\nonumber
\begin{split}
\mathbf{D}^{11}_{k-1}
&={\rm{E}}\{\big(\nabla_{\mathbf{p}_{k-1}}\mathbf{m}(\mathbf{p}_{k-1})\big)\mathbf{Q}^{-1}_{u}\big(\nabla_{\mathbf{p}_{k-1}}\mathbf{m}(\mathbf{p}_{k-1})\big)^T\}\\
&=\begin{bmatrix}  m^2(\sigma_{u}^2)^{-1}   & 0  \\   0   & (\sigma_{m}^2)^{-1}  \end{bmatrix},\\
%
\mathbf{D}^{12}_{k-1}&=-{\rm{E}}\{\nabla_{\mathbf{p}_{k-1}}\mathbf{m}(\mathbf{p}_{k-1})\}\mathbf{Q}^{-1}_{u}\\
&=\begin{bmatrix}  -m(\alpha_{k-1}\sigma_{u}^2)^{-1}   & 0  \\   0   & -(\sigma_{m}^2)^{-1}    \end{bmatrix}.
\end{split}
\end{equation}

The conditional joint probability density function of the observation likelihood distribution is expressed as:
\begin{equation}\nonumber
\begin{split}
p(\mathbf{q}_{k}|\mathbf{p}_{k})=\sum_{j=1}^{N_g}\frac{b^j_{k}}{2\pi\sigma_{T}\alpha_{k}\Lambda^j_{k}}e^{-\frac{s_{k}^2}{2\alpha^2_{k}(\Lambda^j_{k})^2}-\frac{(\tilde{T}_{k}-T_{k})^2}{2\sigma_{T}^2}},
\end{split}
\end{equation}
where $s_k=\alpha_k(t_{k}^{\mathsf{2}}-t_{k-1}^{\mathsf{2}}-t_{k}^{\mathsf{1}}+t_{k-1}^{\mathsf{1}})+\beta_k\tau\big(\kappa(\tilde{T}_k-T_{0})^2+\theta_{0}\big)-\tau\theta_{k}$. The Fisher component $\mathbf{D}^{22}_{k-1}$ is calculated as:
\begin{equation}\nonumber
\begin{split}
\mathbf{D}^{22}_{k-1}
&=\begin{bmatrix}    (\alpha^2_{k-1}\sigma_{u}^2)^{-1}  &  0    \\     0  &  (\sigma_{m}^2)^{-1}   \end{bmatrix}+\frac{1}{\sum_{j=1}^{N_g}(\frac{b^j_{k}}{2\pi\alpha_{k}\Lambda^j_{k}\sigma_{T}})}\\
&\sum_{j=1}^{N_g}\Big(\frac{b^j_{k}}{2\pi\alpha_{k}\Lambda^j_{k}\sigma_{T}}\begin{bmatrix}    \frac{\tau^2}{\alpha^2_{k}(\Lambda^j_{k})^2}  &  0    \\     0  &  \frac{1}{\sigma_{T}^2}   \end{bmatrix}\Big).
\end{split}
\end{equation}

Thus, the Fisher information matrix $\mathbf{J}^{\mathbb{F}}_{k}$ of clock skew estimation based on data fusion is obtained from diagonal matrix forms of Fisher components. 

\footnotesize
\bibliographystyle{IEEEtran}
\bibliography{ref}


\end{document}